\documentclass[10pt,preprint]{aastex}
\usepackage{graphicx,apjfonts,emulateapj5,psfig,onecolfloat5}

\begin{document}
\twocolumn[

\title{Modeling the frequency-dependence of radio beams for cone-dominant pulsars}

\author{P.~F. Wang, 
        J.~L. Han, 
        C. Wang 
       }

\affil{National Astronomical Observatories, Chinese Academy of Sciences,
       A20 Datun Road, Chaoyang District, Beijing 100012, China.
       Email: pfwang, hjl, wangchen@nao.cas.cn
      }

\begin{abstract}
Beam radii for cone-dominant pulsars follow a power-law relation with
frequency, $\vartheta = (\nu/\nu_0)^k+\vartheta_0$, which has not yet
well explained in previous works. We study this frequency dependence
of beam radius (FDB) for cone-dominant pulsars by using the curvature
radiation mechanism. Considering various density and energy
distributions of particles in the pulsar open field line region, we
numerically simulate the emission intensity distribution across
emission height and rotation phase, and get integrated profiles at
different frequencies and obtain the FDB curves. For the density model
of a conal-like distribution, the simulated profiles always shrink to
one component at high frequencies. In the density model with two
separated density patches, the profiles always have two distinct
components, and the power-law indices $k$ are found to be in the range
from $-0.1$ to $-2.5$, consistent with observational results. Energy
distributions of streaming particles have significant influence on the
frequency-dependence behavior. Radial energy decay of particles are
necessary to get proper $\vartheta_0$ in models. We conclude that by
using the curvature radiation mechanism, the observed frequency
dependence of beam radius for the cone-dominant pulsars can only be
explained by the emission model of particles in two density patches
with a Gaussian energy distribution and a radial energy loss.
\end{abstract}


]

\section{Introduction}

A pulsar profile is explained as being the line-of-sight cuts through
the emission beam when a pulsar rotates. For cone-dominant pulsars,
profiles widen at low frequencies. The variation of pulsar profile
width or component separation, $W$, against the observation frequency,
$\nu$, can be described by a power-law function \citep{tho91a}
\begin{equation}
 W=a\nu^k+W_0.
\end{equation}
Here, $W_0$ is a width constant. With a dipole field geometry for
pulsar emission region, the emission of lower frequencies is believed
to be generated at larger heights for the wider open angles of the
emission cone. This is well known as the radius-to-frequency mapping
\citep{rs75,cor78,phi92a}.

Observed profile width can be expressed by, for example, the peak
separation between the outermost components, $W_{pp}$ (i.e. the
peak-peak separation), the pulse width at the $10\%$ of the highest
intensity peak, $W_{10}$, and the pulse width at the $50\%$ of the
peak, $W_{50}$. In order to understand the frequency dependence of
pulsar profiles in a much wider frequency range, \citet{xkj+96}
observed a number of nearby bright pulsars up to 32 GHz and measured
their $W_{50}$. Combining with previous measurements at lower
frequencies, they fitted the pulse widths by $W_{50}=a\nu^k+W_0$, and
got $-0.29>k>-0.94$ for a sample of pulsars.

\begin{table}[h]
\small
\begin{center}
\scriptsize
\caption{Geometry parameters and fitted $k$ and $\vartheta_0$
    for 7 cone-dominant pulsars with clear frequency dependence
    \citep{mr02a}.}
\label{psr-para}
\tabcolsep 1.5mm
\begin{tabular}{ccccccc}
\hline \hline
 PSR & P(s) & $\alpha(^{\circ})$ & $\beta(^{\circ})$ & $\nu_0$(GHz) & $k$ & $\vartheta_0 (^{\circ})$ \\
\hline
 B$0301+19$&$1.388$&$38\pm5$&$1.7\pm0.3$&$2.10\pm0.90$&$-0.47\pm0.06$&$2.3\pm0.2$\\
 B$0329+54$&$0.715$&$32\pm3$&$3.0\pm0.9$&$0.26\pm0.05$&$-1.07\pm0.17$&$6.0\pm0.1$ \\
 B$0525+21$&$3.745$&$21\pm2$&$0.6\pm0.1$&$0.40\pm0.01$&$-0.40\pm0.08$&$1.7\pm0.2$\\
 B$1133+16$&$1.188$&$46\pm3$&$4.1\pm0.5$&$0.12\pm0.03$&$-0.55\pm0.03$&$4.4\pm0.1$\\
 B$1237+25$&$1.382$&$53\pm3$&$0.9\pm0.1$&$1.10\pm0.20$&$-0.45\pm0.05$&$3.0\pm0.2$\\
 B$2020+28$&$0.343$&$56\pm5$&$8.0\pm1.4$&$0.17\pm0.08$&$-0.89\pm0.11$&$8.7\pm0.1$\\
 B$2045-16$&$1.962$&$34\pm2$&$1.1\pm0.1$&$0.40\pm0.05$&$-0.42\pm0.01$&$2.7\pm0.2$\\
\hline
\end{tabular}
\end{center}
\end{table}

The radius of the pulsar emission beam, $\vartheta$, is related
to the profile width, $W$, by \citep{gil81},
\begin{equation}
\vartheta=
2\arcsin[\sin^2\frac{W}{4}\sin\alpha\sin(\alpha+\beta)
+\sin^2\frac{\beta}{2}]^{1/2}.
\label{beam}
\end{equation}
Here, $\alpha$ is the inclination angle of the magnetic axis related
to the rotation axis, $\beta$ is the impact angle of the line of sight
to the magnetic axis. For many pulsars with good polarization
measurements, the values of $\alpha$ and $\beta$ have been determined
\citep{lm88,ran90,ew01}. For a given pulsar, the beam radius,
$\vartheta$, and the profile width, $W$, are quasi-linearly related
for various frequencies. Therefore, the variation of pulse-width with
frequency should be physically related to the frequency dependence of
pulsar beam radius. \citet{mr02a} collected observed pulse widths for
7 cone-dominant pulsars and calculated their beam radii at various
frequencies. They found that the radius of the outer-cone beam of 7
pulsars follows
\begin{equation}
\vartheta =(\nu/\nu_0)^k+\vartheta_0 ,
\label{FDB}
\end{equation}
here $\vartheta_0$ is the beam radius at the infinite-frequency, and
the $\nu_0$ is the characteristic frequency. The values of the
power-law index $k$ are in the range of $-0.3$ to $-1.2$. The values
of $\beta$ are smaller than those of $\vartheta_0$, as shown in
Table~\ref{psr-para}. We noticed that for a given pulsar the $k$ value
obtained from pulse widths is almost the same as that fitted from the
beam radii. \citet{mr02a} noticed that the beam radii of inner-cone do
not show the frequency dependence.

The frequency dependence of beam radii or profile widths can be easily
explained by the open field-lines of the emission region in a dipole
field geometry of neutron stars, as long as the
radius-frequency-mapping holds \citep{rs75}. Different emission
mechanisms can lead to different power-law indices $k$. The electron
bremsstrahlung model \citep{vj73} predicts $k\sim -0.45$. The vacuum
inner gap model \citep{rs75} can give $k\sim -1/3$. The curvature
plasma model \citep{bgi88} can have two extremes for the ordinary
emission mode, either $k\sim -0.14$ or $k\sim-0.29$. The emission from
the cyclotron instability \citep{mu89} gives $k\sim -0.17$. These
theoretical values can not cover the wide range of the observed $k$.

In this paper, we try to explain the frequency dependence of the radio
beam observed for cone-dominant pulsars using the curvature radiation
mechanism. If the edge of pulsar beam is generated by particles
flowing along the last open field-lines (LOF), we can calculate the
radio beams, and investigate their dependence on pulsar parameters,
period $P$, inclination angle $\alpha$, impact angle $\beta$ and the
Lorentz factor of particles $\gamma$. Using some generalized energy
and density distributions of particles in the magnetosphere, we
numerically calculate the radio emission beams and fit their frequency
dependence. We also investigate the influence on the pulsar beam by
the radial decrease of particle energy and the particle energy
distribution.

\section{Curvature radiation beam for particles with a given $\gamma$ in a dipole field}

In general, pulsar radio emission is assumed to be generated by
curvature radiation of secondary particles streaming along the last
open field-lines. In the radio emission region, magnetic fields can be
described as a static dipole \citep[e.g.][]{gan04}, because the
multipolar field components of a neutron star vanish there and the
sweeping effect due to rotation is also negligible according to
\citet{dh04}. Therefore, in this paper, we will use the dipole field
to study the frequency dependence of beam radius (FDB).

\begin{figure}[tb]
  \centering
  \includegraphics[angle=0,width=0.45\textwidth]{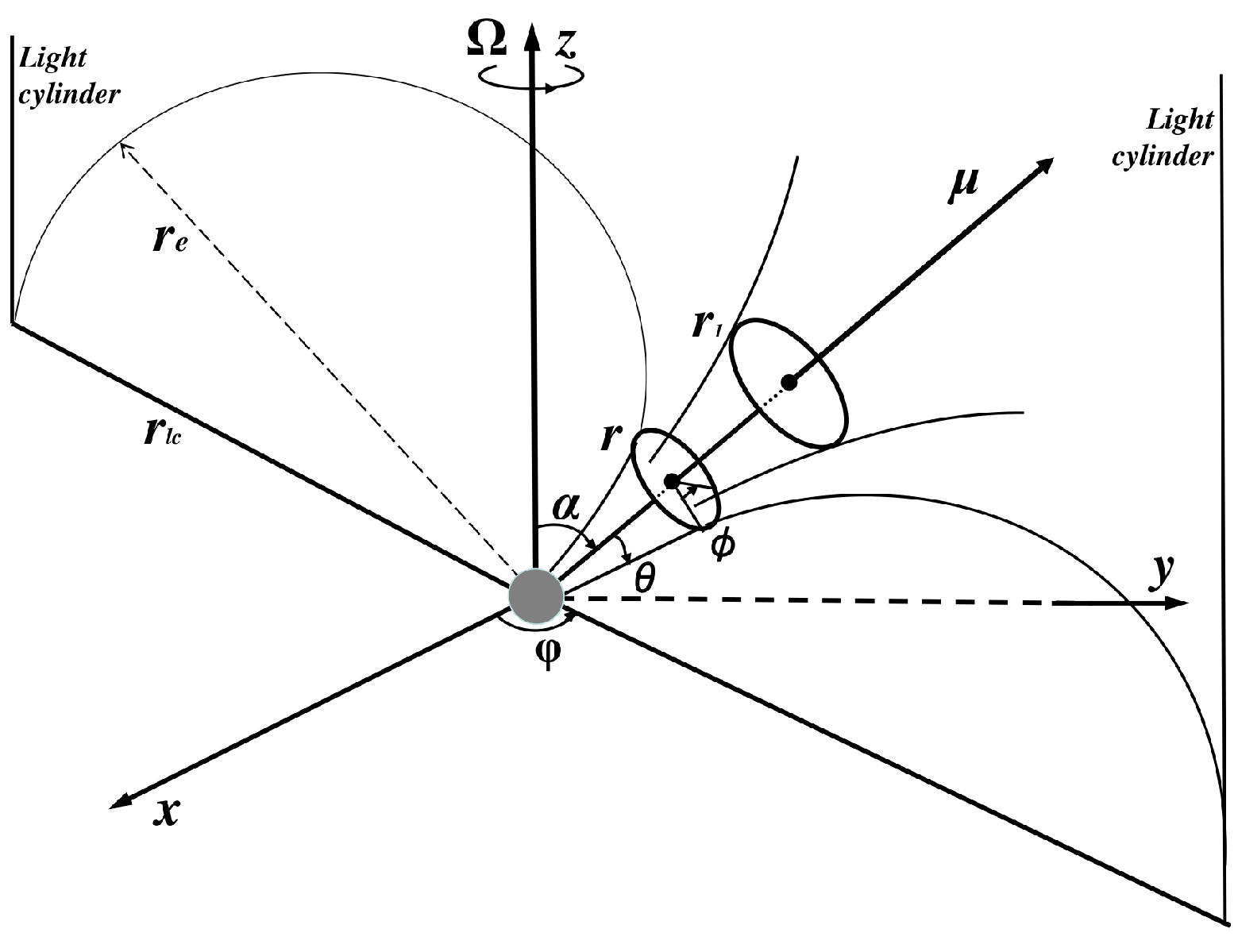}
  \caption{Geometry and parameters defined for an emitting beam.
    $\Omega$ indicates the rotation axis in $z$ direction, $\mu$
    represents the magnetic axis, which is inclined by an angle of
    $\alpha$ with respect to $\Omega$. The LOF is tangential to the
    light cylinder, which has a polar angle $\theta$ from the magnetic
    axis $\mu$.  The radiation beam has a radius of $\vartheta$ at
    a height of $r$.}
\label{geo}
\end{figure}

\begin{figure}[tb]
  \centering
  \includegraphics[angle=0,width=0.40\textwidth]{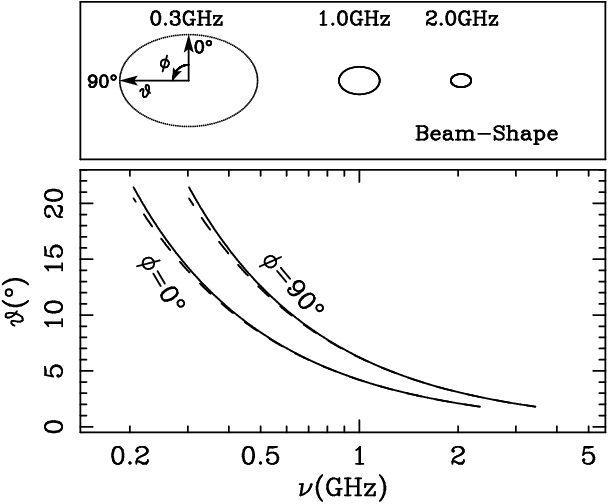}
  \caption{Frequency dependence of the beam radius (FDB curve)
    calculated from the curvature radiation from particles of a single
    $\gamma$ for an inclined dipole of $\alpha=60^{\circ}$. Beam radii
    in the magnetic azimuth of $\phi=0^{\circ}$ and $\phi=90^{\circ}$
    are frequency dependent, though the beam size of $\phi=0^{\circ}$
    is smaller than that of $\phi=90^{\circ}$ at any frequency. This
    behavior is shown in the top panel for 300 MHz, 1.0 GHz and 2.0
    GHz emission. The solid lines are calculated using two terms in
    Eq.~(\ref{appro}), and the dashed lines are only for its first
    term. Here period $P=1$~s and $\gamma=400$ are taken for
    calculations.}
  \label{beam1gma}
\end{figure}

The size and geometry of a dipole magnetic field is determined by
pulsar period $P$, inclination angle $\alpha$ and impact angle
$\beta$. In the polar coordinate system with the polar axis along the
magnetic axis direction (Fig.~\ref{geo}), a dipole field-line can be
described by $r=r_e~\sin^2\theta$, here $\theta$ is the polar angle
from the magnetic axis, $r$ is the distance from the dipole origin,
$r_e$ is the field-line constant, which is the distance from the
origin to the point of the field-line intersection with the magnetic
equatorial plane of $\theta=90\degr$. For an inclined dipole, the LOF
are contained in the light cylinder (see Fig.~\ref{geo}). The radius
of the light-cylinder, $R_{\rm lc}=cP/2\pi$, gives the limit of the
field line constants $r_e$ for the LOF, which is different for pulsars
with different periods. The angular diameter of the polar cap
defined by the feet of LOF on the neutron star surface is related to
the pulsar period, $P$, by $2\theta_{\rm pc}=1.6\degr
P^{-1/2}$. The opening angle of the beam from the tangents of the LOF
near the surface is about $1.2\degr P^{-1/2}$, which defines the
minimum geometrical beam angle. Observations show
$\vartheta_0>1.2\degr P^{-1/2}$. The emission beam determined by the
LOF is not circular but compressed in the meridional direction in the
plane of rotation and magnetic axes \citep{big90b}. We define the
magnetic azimuthal angle, $\phi$, starting from the connection between
the magnetic axis and the rotation axis (to the top being the north)
as being $\phi=0^{\circ}$ (see Fig.~\ref{beam1gma}). The beam radius
in the direction of $\phi=0^{\circ}$ is smaller than that in the
direction of $\phi=90^{\circ}$ (to the east).

In the curvature radiation mechanism, the emission frequency is not
only related to the field geometry but also to the Lorentz factor
$\gamma$ of particles. The simplest case we consider here is that
particles have the same Lorentz factor $\gamma$, and that the radio
beam is defined by the tangents of the LOF. When a relativistic
particle streams along a field-line, it can produce curvature
radiation with a characteristic frequency of
\begin{equation}
\nu=\frac{3\gamma^3c}{4\pi\rho}.
\label{nu}
\end{equation}
Here, $\gamma$ is the Lorentz factor of particles in the range
$10^2-10^4$, $\rho$ is the curvature radius of the particle
trajectory.
In any field-line, the curvature radius can be expressed by \citep{gan04}
\begin{equation}
\rho={r_e}\frac{\sin\theta(5+3\cos2\theta)^{3/2}}{3\sqrt{2}(3+2\cos2\theta)}.
\label{rho}
\end{equation}
Note that $\theta$ varies with $r$. The angle
between the tangent of a field-line of $\theta$ and the magnetic axis
is
\begin{equation}
\vartheta=\arccos(\frac{1+3\cos2\theta}{\sqrt{10+6\cos2\theta}}).
\label{vtheta}
\end{equation}
Combining Eq.~(\ref{rho}) and Eq.~(\ref{vtheta}), one can find the
relation between $\vartheta$ and $\rho$ for any field-line of
$r_e$. Because the radiation frequency $\nu$ is related to $\rho$ by
Eq.~(\ref{nu}), we get
\begin{equation}
\vartheta=15.4^\circ(\frac{\gamma^3c}{r_e\nu})+
0.43^\circ(\frac{\gamma^3c}{r_e\nu})^{3}.
\label{appro}
\end{equation}
One can use Eq.~(15) from \citet{gan04} to get the field-line
constants $r_e$ for each LOF.
Obviously, as indicated in Eq.~(\ref{appro}), the beam radius is
related to the radiation frequency. At a higher frequency of $\nu \gg
\gamma^3c/r_e$, the second term can be neglected, and the beam radius
$\vartheta$ is related to the frequency $\nu$ by roughly a power-law
with the index of $k=-1$. At a low frequency of $\nu \leq \gamma^3c/r_e$,
the frequency dependence becomes slightly steeper due to the
contribution from the second term. Note that the beam radii in all
magnetic azimuthal directions are frequency dependent. Clearly, the
first term of Eq.~(\ref{appro}) is a good approximation of the
frequency dependence of pulsar beam, which has a power-law index
$k=-1$.

\begin{figure}[tb]
  \centering
  \includegraphics[angle=0,width=0.45\textwidth]{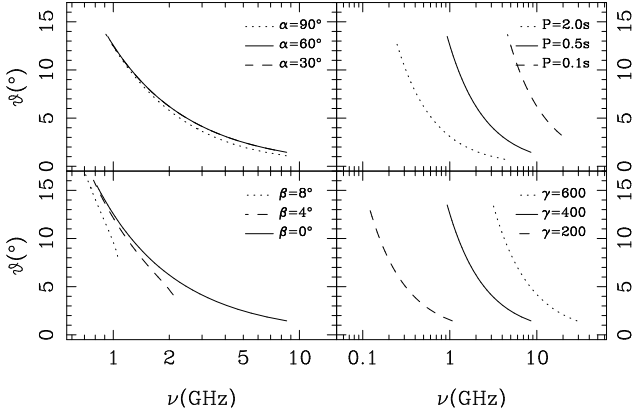}
  \caption{Beam radii at a series of frequencies calculated from the
    curvature radiation of particles with a single $\gamma$. Default
    model parameters are $P=0.5$~s, $\alpha=60^{\circ}$,
    $\beta=0^{\circ}$ and $\gamma=400$. Various curves have been
    calculated for different parameters, $\alpha$, $\beta$, $P$ and
    $\gamma$ in four panels. Note that the two curves for
    $\alpha=60\degr$ and $\alpha=30\degr$ in the upper left panel are
    completely overlapped. }
  \label{cb4para}
\end{figure}

We calculated the beam radii at different frequencies for various
model parameters, as shown in Fig.~\ref{cb4para}. All curves show the
frequency dependence of radiation beam with a power-law index of
approximately $-1$, which is different from observational values in
the range of $-0.3>k>-1.2$ \citep{mr02a}.  We found that there is no
influence on the frequency dependence of beam size by the magnetic
inclination $\alpha$, because $\alpha$ only leads to the compression
of beam in the meridional direction, and has almost no influence on
the radius of $\phi=90^{\circ}$. Different impacts of line of sight
with different $\beta$ values lead to different cuts of the beam from
$\beta=0^{\circ}$ to $\beta=\vartheta$.  The lower limit of
$\vartheta$ is determined by the angular size of the polar cap
(3/2$\theta_{\rm pc}$) or the $\beta$ value. Pulsars with a small
period $P$ have a small $R_{\rm lc}$ and hence a small $r_e$ and a
small curvature radius [Eq.~(\ref{rho})], which corresponds to a
larger radiation frequency [Eq.~(\ref{nu})], as shown in
Fig.~\ref{cb4para}. Note also that in the pulsar emission region
curvature radiation of particles with a single $\gamma$ cannot produce
radio emission over a wide frequency range from hundreds MHz to ten
GHz. Particles with larger $\gamma$ produce curvature radiation at
much higher frequency, which shifts the FDB curves to higher frequency
ranges.

The calculations shown in Fig.\ref{beam1gma} and \ref{cb4para} were
made with assumptions that pulsar beams are bounded by the LOF.
However, \citet{rs75} suggested that beam edge should be bounded by
the critical field-lines, which are orthogonal to (instead of
tangent to) the light cylinder at the intersection points. The
critical field-lines are located between the magnetic axis and the
LOF. Here, the parameter $\eta$ is used to describe the location of
field-lines, with $\eta=0$ for the magnetic axis, $\eta=1$ for the
LOF. To generate the curvature emission beam of the same open angle,
the emission height from the critical field lines ($0<\eta<0.74$,
depending on $\alpha$) is larger than that from the LOF as shown in
Fig.~\ref{cb4para}, and the curvature radius is also larger, so that
the emission has a smaller frequency of,
\begin{equation}
\nu=\nu_{\rm lof}\times \eta^2=\frac{3\gamma^3c}{4\pi\rho} \eta^2,
\end{equation}
with $\nu_{\rm lof}$ the corresponding emission frequency for the LOF.
We carried out a set of calculations, and found that the curvature
radiation for the critical field lines has almost the same frequency
dependence of emission beam as that for the LOF, and $k$ and
$\vartheta_0$ values are consistent within 5\%, though $\nu_0$ is
smaller due to larger curvature radii of field lines.

\section{Curvature radiation beam from particles with various energy and density distributions}

Particles in pulsar magnetosphere should have an energy distribution,
which radiate in a range of frequencies at a range of heights for a
given LOF. Furthermore, particles flow out along a set of open
field-lines, rather than just the LOF. The pulsar radio emission from
a given height and a given rotation phase is contributed from
particles not only in the field lines which are tangential towards the
observer, but also in the nearby field lines in the bunch within the
$1/\gamma$ emission cone. The emission is coherent radiation
from a bunch of particles \citep{bb76}.

According to simulation results of \citet{ml10}, we assume in this
section that secondary particles for curvature radiation at radio
bands follow a Gaussian energy distribution with a peak at $\gamma_m$
of several hundreds:
\begin{equation}
n_{\rm e}(\gamma)\sim \exp[-\frac{(\gamma-\gamma_m)^2}{2\sigma_\gamma^2}].
\label{gammadis}
\end{equation}
Here, the standard deviation $\sigma_\gamma$ is of several
tens. Considering the continuity of the particles flowing along the
field-line tube, we got the number density of particles at $r$
as
\begin{equation}
n_{\rm e}(r)=n_{\rm e0} (r/R_\star)^{-3}.
\end{equation}
Here, $n_{\rm e0}$ represents the number density at the bottom of a
magnetic field tube near the surface of a neutron star.

The power of curvature radiation at a frequency
$\nu$ from one particle is given by,
\begin{equation}
P_{\rm e}=\frac{2q^2\gamma^4}{3c}(\frac{c}{\rho})^2.
\label{power}
\end{equation}
$N_b$ particles in a field bunch in the region with a dimension size
of less than half emission wavelength produce the total emission power
of approximately $N_b^2\;P_{\rm e}$.

Because the curvature radius varies everywhere in the dipole field,
according to Eq.~(\ref{nu}) and (\ref{power}), the observed emission
at frequency $\nu$ and rotation phase $\varphi$ should come from the
tangents of a set of open field-lines with the same magnetic azimuth
$\phi$, where the curvature radius $\rho$ of a field-line and particle
energy $\gamma$ are nicely combined to produce emission. The observed
total power therefore should be the sum of them,
\begin{equation}
  I(\nu,\varphi)=\int_{r_{\min}}^{r_{\max}}n_{\rm e}(r,\gamma)N_b^2
  P_{\rm e}(r,\phi,\gamma)dr .
  \label{rinten}
\end{equation}
Note that the magnetic azimuth $\phi$ is a monotonic function of
the rotation phase $\varphi$, as given by Eq.~(11) in \citet{gan04}
which is equivalent to the well-known $S$-curve for polarization
angle.

\begin{figure}[tb]
\centering
\includegraphics[angle=0, width=0.25\textwidth]{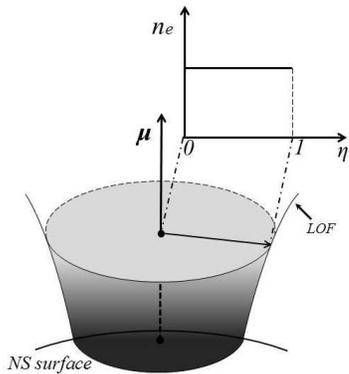}
\caption{Illustration for the uniform density distribution of
  particles in pulsar magnetosphere. Here, $n_{\rm e}=n_{\rm e0}
  (r/R_\star)^{-3}$.}
\label{fig:umodel}
\end{figure}

\begin{figure}[tb]
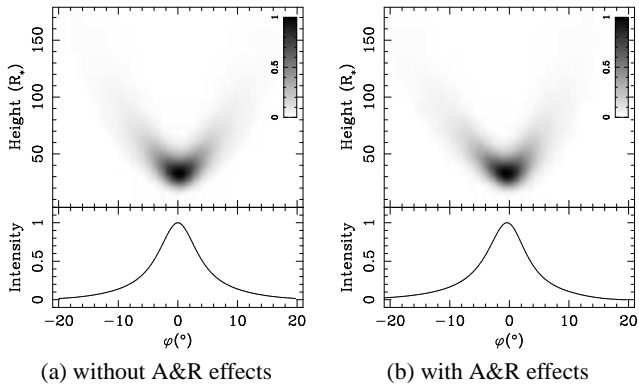

\centering
\begin{tabular}{cc}
  \includegraphics[angle=0, width=0.215\textwidth]{U-image200-NoAR.ps} &
  \includegraphics[angle=0, width=0.215\textwidth]{U-image200-AR.ps} \\
  {\small (a) without A\&R effects}   & {\small (b) with A\&R effects} \\
\end{tabular}
\caption{{\it (a):} {\it Upper panel:} Emission intensities at a given
  frequency (here $\nu=200$~MHz) for various rotation phases $\varphi$
  and emission heights, radiated by a group of particles which have a
  Gaussian energy distribution and an uniform number distribution in
  magnetic polar angles as shown in Fig.~\ref{fig:umodel}. {\it Lower
    panel:} The accumulated emission profile (also at
  $\nu=200$~MHz). The model parameters used are $P=1$s,
  $\alpha=45^{\circ}$ and $\beta=3^{\circ}$, and $\gamma_m=350$ and
  $\sigma_\gamma=30$.  {\it (b):} Same as left panels {\it (a)}, but
  the abberation and retardation effects have been considered.}
\label{uimage}
\end{figure}

\subsection{Emission beam from particles uniformly distributed
in magnetic polar angles}

As the first try, we assume that particles follow a Gaussian energy
distribution and are uniformly distributed in the open field-lines for
all magnetic polar angles, as shown in Fig.~\ref{fig:umodel}.

We can calculate the emission from each open field-line at every
tangent for each frequency. At a given emission height, the central
open field-lines have larger curvature radii than the outer open
field-lines. Particles with a given Lorentz factor in the central
field lines would produce a lower frequency emission [see
  Eq.~(\ref{nu})] than they do in the outer field lines. For a group
of particles with a Gaussian energy distribution, only these particles
with larger Lorentz factors in the central field lines can produce the
emission at the same frequency as particles of lower Lorentz factors
in the outer field lines, but their emission intensity is reduced by a
factor of $1/\gamma^2$. Note that magnetic fields have smaller
curvature radii at lower height. All these factors produce the
bifurcation feature shown in Fig.\ref{uimage}.

We limited our calculations for the region of $r_{\max}< 0.1 R_{\rm
  lc}$, because the emission intensity from higher regions is
negligible (see Fig.~\ref{uimage}) due to the relatively small
particle density and large curvature radius. For $P=1$~s, $R_{\rm
  lc}\simeq 47771$~km $\simeq 4777 R_*$, if we take $R_*=$10~km.  We
consider the phase shift \citep{gan05} of the emission from any height
of every field-line, and integrate their emission for profiles at
every observation frequency (lower panels in Fig.~\ref{uimage}). We
found that aberration and retardation effects can change the
projection position of emission in the sky plane, which cause the
profile to be enhanced by about 20\% at the phase of
$\varphi=-10^\circ$ and weakened by about 22\% at the phase of
$\varphi=10^\circ$ for the model calculations in Fig.~\ref{uimage}.

We conclude that curvature radiation from particles uniformly
distributed in magnetic polar angles cannot produce double conal
profiles.

\begin{figure}[tb]
\centering
\begin{tabular}{cc}
  \includegraphics[angle=0, width=0.25\textwidth]{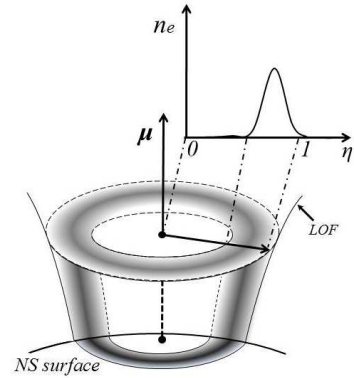}
\end{tabular}
\caption{Illustration for the conal density distribution of particles
  in pulsar magnetosphere. The density follows Eq.~(\ref{cmodel}).}
\label{fig:cmodel}
\end{figure}

\subsection{Emission beam from particles in the conal density distribution}

Small curvature radii of the magnetic field lines near the edge of the
pulsar polar cap make curvature radiation more efficiently being
generated, which then causes the sparkings more probably located
around the edge of polar cap \citep{rs75,qlz+01}. It is therefore
possible that accelerated particles are more likely distributed in a
conal area, instead of an uniform distribution. Noticed that charged
particles flow out along a fixed line tube, we define the density
distribution in a conal area on the polar cap as
\begin{equation}
n_{\rm e0}(\eta)=n_{\rm e0}{\exp}[-\frac{(\eta-\eta_m)^2}{2\sigma_\eta^2}],
\label{cmodel1}
\end{equation}
here $\eta_m$ denotes the peak position of the cone, and the standard
derivation $\sigma_\eta$ describes the cone width. At any radius $r$
and any magnetic polar angle $\eta$ the density is
\begin{equation}
n_{\rm e}(r,\eta)=n_{\rm e0}(\eta) (r/R_\star)^{-3},
\label{cmodel}
\end{equation}
which does not vary with magnetic azimuthal angle $\phi$.
The illustration for the conal density distribution is given in
Fig.~\ref{fig:cmodel}.

\begin{figure}[tb]
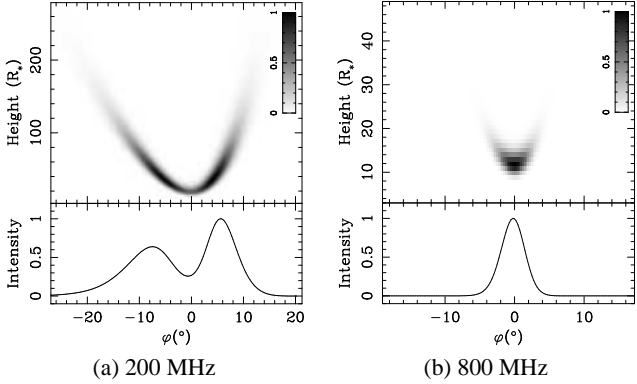

\centering
\begin{tabular}{cc}
  \includegraphics[angle=0, width=0.215\textwidth]{C-image200.ps} &
  \includegraphics[angle=0, width=0.215\textwidth]{C-image800.ps} \\
{\small (a) $200$~MHz}   & {\small (b) $800$~MHz} \\
\end{tabular}
\caption{Same as Fig.~\ref{uimage} for emission intensity distributions
  and the accumulated profiles {\it (a):} at $200$~MHz and {\it (b):}
  at $800$~MHz for a conal density model. The particle energy and
  density distribution parameters are $\gamma_m=350$,
  $\sigma_\gamma=30$, $\eta_m=0.75$ and $\sigma_\eta=0.08$, with
  $P=1$s, $\alpha=45^{\circ}$ and $\beta=5^{\circ}$.}
\label{cimage}
\end{figure}

\begin{figure}[tb]
    \centering
    \includegraphics[angle=0, width=0.22\textwidth]{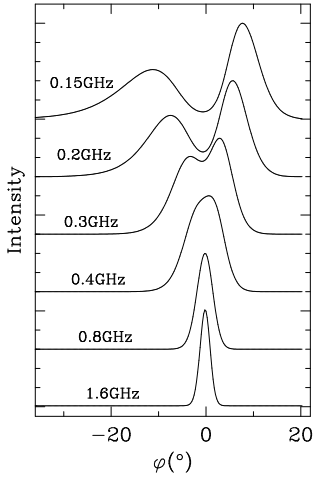}
    \caption{Integrated profiles at a series of frequencies calculated
      from curvature radiation of particles with a conal density
      distribution. The model parameters are the same as those in
      Fig.~\ref{cimage}. The profiles are aligned to the absolute zero
      rotation phase when the sight line is in the meridional plane
      defined by the rotation and magnetic axes.}
    \label{cprofiles}
\end{figure}

Similar to the uniform model in Sect.3.1, we calculated the curvature
radiation of particles in such a density cone from each field line at
every tangent at each frequency, to composite the emission beam
between 100\,MHz to 30\,GHz. Fig.~\ref{cimage} shows the distribution
of emission intensities for various heights and rotation phases as
well as integrated profiles at two example frequencies, 200\,MHz and
800\,MHz. There are two components in the integrated profile at
200\,MHz, though the emission for central rotation phases coming from
the region of several tens of $R_\star$ is relatively strong compared
to those from higher regions. The aberration and retardation effects,
which shift high-altitude emission to early rotation phases, can cause
the leading profile component to be wider and weaker than the trailing
one at 200\,MHz, as demonstrated by \citet{dwd10}. At 800~MHz emission
mostly comes from a lower region of about $r=10\sim 20 R_\star$,
rather than a few hundred of $R_\star$ for 200~MHz emission, where the
sight line cuts only the beam edge and the emission from higher
regions is too weak, so that only one component appears in the
integrated profile. The profile evolves from two components at low
frequencies to one component at frequencies of above 400\,MHz (see
Fig.~\ref{cprofiles}).

\begin{figure}[tb]
    \centering
    \includegraphics[angle=0,width=0.42\textwidth]{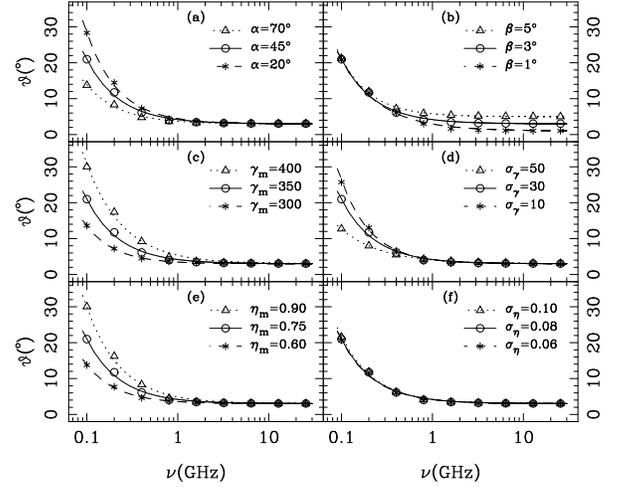}
    \caption{Various FDB curves calculated from particles with a
      Gaussian energy distribution and a conal density distribution
      above the polar cap with different model parameters of $\alpha$,
      $\beta$, $\gamma_m$, $\sigma_\gamma$, $\eta_m$ and
      $\sigma_\eta$. The default parameters are the same as those in
      Fig.~\ref{cimage}.}
    \label{crelation}
\end{figure}
\begin{table}[tb]
\small
\begin{center}
\scriptsize
\caption{Fitted parameters for various FDB curves in
  Fig.~\ref{crelation} from various conal density models with different
  parameters of $\alpha$, $\beta$, $\gamma_m$, $\sigma_\gamma$,
  $\eta_m$ and $\sigma_\eta$}
\label{tcfdb}
\tabcolsep 1.0mm
\begin{tabular}{cccccc|cccc}
\hline
\hline
 $\alpha$ &$\beta$&$\gamma_m$&$\sigma_\gamma$&$\eta_m$&$\sigma_\eta$&$\nu_0 (\rm GHz)$&$k$& $\vartheta_0 (^{\circ})$& $\vartheta_{\rm 100 Mhz} (^{\circ})$\\
\hline
 $20^{\circ}$&$3^{\circ}$&$350$&$30$&$0.75$&$0.08$&$1.35\pm0.10$&$-1.24\pm0.04$&$3.00$&$28.55\pm0.38$  \\
 $45^{\circ}$&     -   &   -  & -  &  -  &   -   &$1.15\pm0.12$&$-1.18\pm0.05$&$3.00$&$21.20\pm0.39$  \\
 $70^{\circ}$&     -   &   -  & -  &  -  &   -   &$0.75\pm0.07$&$-1.18\pm0.06$&$3.00$&$13.89\pm0.26$  \\
 \hline
 $45^{\circ}$&$1^{\circ}$&$350$&$30$&$0.75$&$0.08$&$1.93\pm0.21$&$-1.01\pm0.04$&$1.00$&$21.15\pm0.40$  \\
     -     &$3^{\circ}$&  -   & -  &  -  &  -    &$1.15\pm0.12$&$-1.18\pm0.05$&$3.00$&$21.20\pm0.39$  \\
     -     &$5^{\circ}$&  -   & -  &  -  &  -    &$0.84\pm0.05$&$-1.31\pm0.04$&$5.00$&$21.33\pm0.24$  \\
\hline
 $45^{\circ}$&$3^{\circ}$&$300$&$30$&$0.75$&$0.08$&$0.57\pm0.01$&$-1.35\pm0.02$&$3.00$&$13.63\pm0.07$  \\
     -     &     -    &$350$&  - &  -   &  -   &$1.15\pm0.12$&$-1.18\pm0.05$&$3.00$&$21.20\pm0.39$  \\
     -     &     -    &$400$&  - &  -   &  -   &$2.23\pm0.34$&$-1.07\pm0.06$&$3.00$&$30.46\pm0.72$  \\
\hline
 $45^{\circ}$&$3^{\circ}$&$350$&$10$&$0.75$&$0.08$&$1.12\pm0.11$&$-1.30\pm0.06$&$3.00$&$25.94\pm0.56$  \\
     -     &     -    &  -  &$30$&  -   &  -   &$1.15\pm0.12$&$-1.18\pm0.05$&$3.00$&$21.20\pm0.39$  \\
     -     &     -    &  -  &$50$&  -   &  -   &$1.06\pm0.02$&$-0.96\pm0.01$&$3.00$&$12.74\pm0.05$  \\
\hline
$45^{\circ}$&$3^{\circ}$&$350$&$30$&$0.60$&$0.08$&$0.65\pm0.03$&$-1.27\pm0.03$&$3.00$&$13.83\pm0.13$ \\
     -     &     -    &  -  &  - &$0.75$&  -   &$1.15\pm0.12$&$-1.18\pm0.05$&$3.00$&$21.20\pm0.39$  \\
     -     &     -    &  -  &  - &$0.90$&  -   &$1.74\pm0.20$&$-1.15\pm0.05$&$3.00$&$30.34\pm0.58$ \\
\hline
$45^{\circ}$&$3^{\circ}$&$350$&$30$&$0.75$&$0.06$&$1.12\pm0.12$&$-1.19\pm0.06$&$3.00$&$20.89\pm0.41$ \\
     -     &     -    &  -  &  - &  -   &$0.08$&$1.15\pm0.12$&$-1.18\pm0.05$&$3.00$&$21.20\pm0.39$  \\
     -     &     -    &  -  &  - &  -   &$0.10$&$1.17\pm0.10$&$-1.19\pm0.04$&$3.00$&$21.74\pm0.34$ \\
\hline
\end{tabular}
\end{center}
\end{table}

We measured the profile width at 10\% of the peak intensity, and
calculate the corresponding beam radius $\vartheta$ using
Eq.~(\ref{beam}). The FDB curves are plotted in Fig.~\ref{crelation}
for various initial conditions. These curves are fitted using
Eq.~(\ref{FDB}), and results are listed in Table~\ref{tcfdb}.  The
profile width is close to zero when $\beta=\vartheta$. We have to fix
$\vartheta_0$ to be the same as the impact angle $\beta$ during the
fitting, because the emission at very high frequencies comes from the
lowest emission point on the LOF. As seen in Fig.~\ref{crelation} and
Table.~\ref{tcfdb}, the FDB curves are closely related to emission
geometry ($\alpha$, $\beta$), particle energy ($\gamma$,
$\sigma_\gamma$) and conal density distributions ($\eta_m$).

We noticed that the FDB curves in all panels of Fig.~\ref{crelation}
are converged to a similarly flattened FDB curve at high frequencies,
except for various $\beta$ in panel {\it b}. As leaned from
Fig.~\ref{cimage}, the emission at high frequency comes from a very
small height, so that the emission beam is rather narrow due to the
geometrical effect. The density distribution ($\eta_m$, $\sigma_\eta$)
and energy distribution ($\gamma$, $\sigma_\gamma$) of particles as
well as the magnetic inclination ($\alpha$) do not have any obvious
effect on the beam at high frequencies, as shown by the similarly
converged flat FDB curves in all panels in Fig.~\ref{crelation}. As
expected, a sight line with a small impact angle $\beta$ can look into
the much deeper magnetosphere and detect a much smaller beam at high
frequencies (panel {\it b} in Fig.~\ref{crelation}).

The FDB curves in Fig.~\ref{crelation} are influenced by $\alpha$,
$\gamma$, $\sigma_\gamma$ and $\eta_m$ more obviously at the lower
frequency end.
For a larger magnetic inclination angle $\alpha$, the polar cap as
well as pulsar beam will be more compressed in the meridian dimension
(Fig.~\ref{beam1gma}). When other model parameters are fixed, for a
dipole field geometry with a larger $\alpha$, the fixed $\eta_m$ means
that the bunch of field lines closer to the magnetic axis are used to
calculate the emission beam, which have large curvature radii and
emission frequencies are systematically smaller. The characteristic
frequency $\nu_0$ of the FDB curves is therefore also smaller (see
Table~\ref{tcfdb}).
Particles with a larger $\gamma_m$ produce emission of a given
frequency at higher altitudes with larger curvature radii, which
produce a wider beams (panel {\it  c} in Fig.~\ref{crelation}) with
a large characteristic frequency $\nu_0$ (Table~\ref{tcfdb}).
If the energy distribution of particles is rather wide, e.g.  for a
model with a larger $\sigma_\gamma$, the emission region at a given
frequency is very extended along field lines. However the dominant
emission always comes from a lower emission height, and the integrated
profile and hence the beam is much less sensitive to frequencies
(panel {\it d} in Fig.~\ref{crelation}).
The particle distribution ($\eta_m$) is directly related to the field
lines where the emission is produced. It is easy to understand that
particle distributed with a small $\eta_m$ tends to emit with a
smaller beam at any frequencies (panel {\it e} in
Fig.~\ref{crelation}). We noticed that the particle distribution width
$\sigma_\eta$ does not have obvious influence on the FDB curves (
panel {\it f} in Fig.~\ref{crelation}).

\begin{figure}[tb]
\centering
\includegraphics[angle=-90, width=0.44\textwidth] {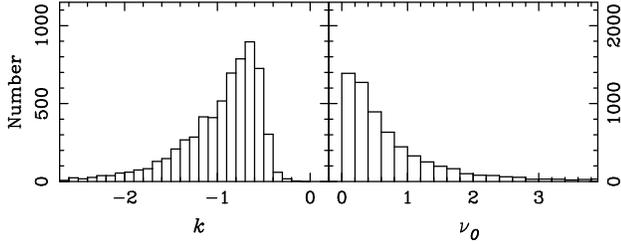}
\caption{Distributions of $k$ and $\nu_0$ from 6375 combinations of
various model parameter values of the conal density model.}
\label{knudis}
\end{figure}

\begin{figure}[tb]
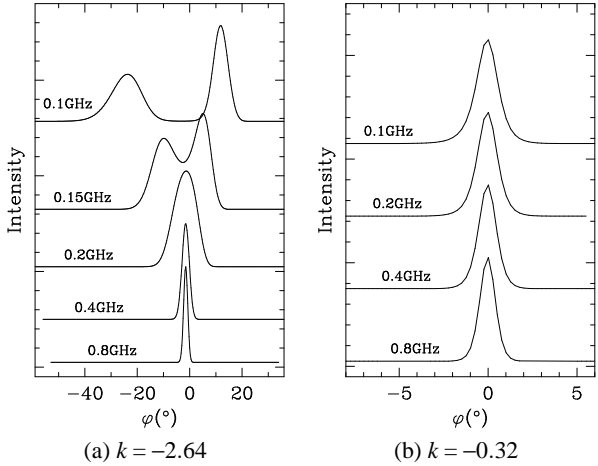

\centering
\begin{tabular}{cc}
\includegraphics[angle=0, width=0.2\textwidth] {C-profiles_largek.ps} &
\includegraphics[angle=0, width=0.2\textwidth] {C-profiles_smallk.ps} \\
{\small (a) $k=-2.64$}  & {\small (b) $k=-0.32$}
\end{tabular}
\caption{Profile evolution with frequency for two extreme $k$ values:
  {\it Panel (a)} for an extreme steep frequency dependence of
  $k=-2.64$ with model parameters of $\alpha=45^\circ$,
  $\beta=9^\circ$, $\gamma_m=300$, $\sigma_\gamma=10$, $\eta_m=0.9$
  and $\sigma_\eta=0.05$; and {\it Panel (b)} for an extreme flat
  frequency dependence of $k=-0.32$ with model parameters of
  $\alpha=65^\circ$, $\beta=1^\circ$, $\gamma_m=200$,
  $\sigma_\gamma=50$, $\eta_m=0.3$ and $\sigma_\eta=0.25$.}
\label{cprof2}
\end{figure}

To further investigate the frequency dependence of pulsar beam radius
on model parameters in a wide range of values, we have calculated the
emission intensity distribution and integrated profiles (as shown in
Fig.~\ref{cimage} and Fig.~\ref{cprofiles}) for 6375 combinations of
different parameter values: $\alpha = (5^\circ$, $25^\circ$,
$45^\circ$, $65^\circ$, $85^\circ$), $\beta = (1^\circ$, $3^\circ$,
$5^\circ$, $7^\circ$, $9^\circ$), $\gamma_m = (200$, $250$, $300$,
$250$, $400$), $\sigma_\gamma = (10$, $30$, $50$), $\eta_m = (0.3$,
$0.5$, $0.7$, $0.9$) and $\sigma_\eta = (0.05$, $0.1$, $0.15$, $0.2$,
$0.25$). Pulsar period is fixed to be $P=1$s. The frequency dependence
of beam radii are fitted with Eq.~(\ref{FDB}) to get $k$ and
$\nu_0$. We got the distributions of $k$ and $\nu_0$ from these
combinations as shown in Fig.~\ref{knudis}. We found that values of
$k$ are in the range of $-0.2$ to $-2.7$, consistent with observed
values. However, profile components always merge to one component
above $\sim300MHz$, which is very different from observational facts
of cone-dominant pulsars \citep{mr02a}. We show in Fig.~\ref{cprof2}
the profile evolutions for two extremes of $k$ values: $k=-2.64$ and
$k=-0.32$. For the case of $k=-0.32$, the two profile components merge
even at a very low frequency below $100MHz$. The merging happens when
$\gamma_m<250$ or $\sigma_\gamma>30$ or $\eta_m<0.5$ or
$\sigma_\eta>0.15$. We noticed that when $\gamma_m>350$, for models of
curvature radiation from particles in the conal density distribution,
the emission below about $100MHz$ comes from the region of several
thousands of $R_\star$, which leads profiles as wide as about
$200^{\circ}$.

In summary, though curvature radiation models of particles in a
conal density distribution can produce a proper range of $k$, the
merging of profile components above few hundreds MHz is the main
reason to reject the models for cone-dominated profiles.

\begin{figure}[tb]
\centering
\includegraphics[angle=0, width=0.25\textwidth]{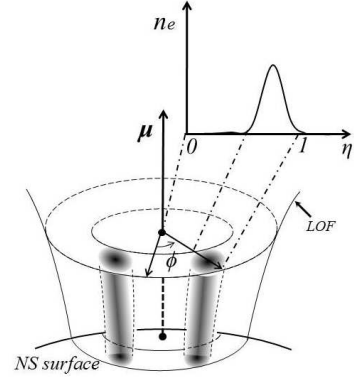}
\caption{Illustration for the particle distribution in two
  density patches. The density follows Eq.~(\ref{pmodel}).}
\label{fig:pmodel}
\end{figure}

\subsection{Emission beam from particles of two density patches}

In the inner gap model \citep{rs75}, the sparkings do not
continually happen above the polar cap and are not concentrating in
a ring area to produce the conal density distribution discussed
above. The most probably case is that the sparkings occur separately
above the polar cap, which produce the subbeams of the observed
conal beam \citep{ran83} or the patchy beam \citep{lm88,hm01}. In
the following we work on the models of curvature radiation of
particles in two density patches which could be reasonably produced
by two sparkings in the inner gap of a rotating neutron star.

\begin{figure}[tb]
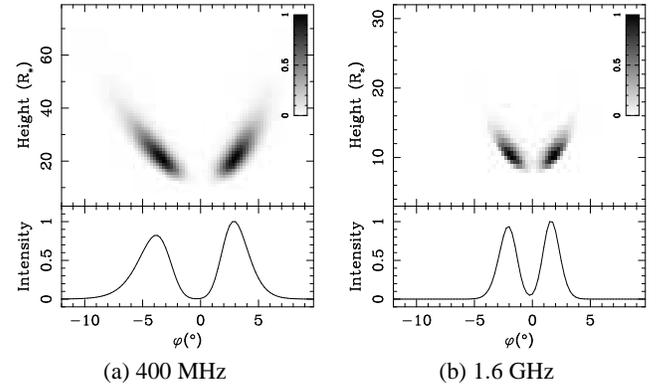

\centering
\begin{tabular}{cc}
  \includegraphics[angle=0,width=0.215\textwidth]{P-image400.ps} &
  \includegraphics[angle=0,width=0.215\textwidth]{P-image1600.ps} \\
  {\small (a) $400$~MHz}  & {\small (b) $1.6$~GHz}  \\
\end{tabular}
\caption{Same as Fig.~\ref{uimage} for emission intensity
  distributions and the accumulated profiles {\it (a):} at $400$~MHz
  and {\it (b):} at $1.6$~GHz for a model of two density patches. The
  particle energy and density distribution parameters are
  $\gamma_m=350$, $\sigma_\gamma=30$, $\eta_m=0.75$,
  $\sigma_\eta=0.08$, $\phi_m=45^{\circ}$ and
  $\sigma_\phi=12.5^{\circ}$, with $P=1$s, $\alpha=45^{\circ}$ and
  $\beta=3^{\circ}$.}
\label{pimage}
\end{figure}

\begin{figure}[tb]
    \centering
    \includegraphics[angle=0,width=0.22\textwidth]{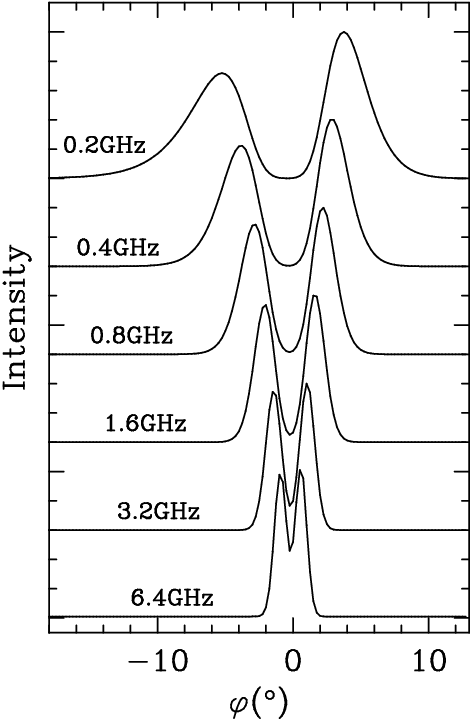}
    \caption{Same as Fig.~\ref{cprofiles} except for the model of two
      density patches. The model parameters are the same as those in
      Fig.~\ref{pimage}.}
    \label{pprofiles}
\end{figure}

\begin{table*}[htb]
\small
\begin{center}
\scriptsize
\caption{Parameters fitted to the FDB curves in Fig.~\ref{prelation}
  for models of two density patches with various $\alpha$, $\beta$,
  $\gamma_m$, $\sigma_\gamma$, $\eta_m$, $\sigma_\eta$, $\phi_m$ and
  $\sigma_\phi$}
\label{tpfdb}
\begin{tabular}{cccccccc|cccr}
\hline
\hline
 $\alpha$ &$\beta$&$\gamma_m$&$\sigma_\gamma$&$\eta_m$&$\sigma_\eta$&$\phi_m$&$\sigma_\phi$ &$\nu_0 (\rm GHz)$&$k$& $\vartheta_0 (^{\circ})$& $\vartheta_{100} (^{\circ})$ \\
 \hline
 $20^{\circ}$&$3^{\circ}$&$350$&$30$&$0.75$&$0.08$&$45^{\circ}$&$12.5^{\circ}$&$1.78\pm0.11$&$-0.66\pm0.02$&$3.00$&$9.71\pm0.10$\\
 $45^{\circ}$&     -   &   -  & -  &  -  &   -   &      -   &     -      &$1.38\pm0.04$&$-0.80\pm0.01$&$3.00$& $11.17\pm0.06$\\
 $70^{\circ}$&     -   &   -  & -  &  -  &   -   &      -   &     -      &$1.03\pm0.03$&$-0.78\pm0.01$&$3.00$& $9.24\pm0.05$\\
 \hline
 $45^{\circ}$&$1^{\circ}$&$350$&$30$&$0.75$&$0.08$&$45^{\circ}$&$12.5^{\circ}$&$1.07\pm0.05$&$-1.08\pm0.02$&$1.00$& $13.84\pm0.14$\\
     -     &$3^{\circ}$&  -   & -  &  -  &  -    &      -   &     -      &$1.38\pm0.04$&$-0.80\pm0.01$&$3.00$& $11.17\pm0.06$\\
     -     &$5^{\circ}$&  -   & -  &  -  &  -    &      -   &     -      &$1.78\pm0.13$&$-0.71\pm0.02$&$5.00$& $12.65\pm0.13$\\
\hline
 $45^{\circ}$&$3^{\circ}$&$300$&$30$&$0.75$&$0.08$&$45^{\circ}$&$12.5^{\circ}$&$1.07\pm0.05$&$-0.66\pm0.02$&$3.00$& $7.78\pm0.07$\\
     -     &     -    &$350$&  - &  -   &  -   &      -    &     -      &$1.38\pm0.04$&$-0.80\pm0.01$&$3.00$& $11.17\pm0.06$\\
     -     &     -    &$400$&  - &  -   &  -   &      -    &     -      &$1.82\pm0.04$&$-0.91\pm0.01$&$3.00$& $16.99\pm0.06$\\
\hline
 $45^{\circ}$&$3^{\circ}$&$350$&$10$&$0.75$&$0.08$&$45^{\circ}$&$12.5^{\circ}$&$1.01\pm0.06$&$-1.18\pm0.03$&$3.00$& $18.49\pm0.20$\\
     -     &     -    &  -  &$30$&  -   &  -   &      -    &     -      &$1.38\pm0.04$&$-0.80\pm0.01$&$3.00$& $11.17\pm0.06$\\
     -     &     -    &  -  &$50$&  -   &  -   &      -    &     -      &$3.20\pm0.69$&$-0.41\pm0.03$&$3.00$& $7.08\pm0.18$\\
\hline
$45^{\circ}$&$3^{\circ}$&$350$&$30$&$0.60$&$0.08$&$45^{\circ}$ &$12.5^{\circ}$&$1.16\pm0.05$&$-0.66\pm0.01$&$3.00$& $8.01\pm0.05$\\
     -     &     -    &  -  &  - &$0.75$&  -   &      -    &     -      &$1.38\pm0.04$&$-0.80\pm0.01$&$3.00$& $11.17\pm0.06$\\
     -     &     -    &  -  &  - &$0.90$&  -   &      -    &     -      &$1.52\pm0.03$&$-0.95\pm0.01$&$3.00$& $16.26\pm0.06$\\
\hline
$45^{\circ}$&$3^{\circ}$&$350$&$30$&$0.75$&$0.06$&$45^{\circ}$ &$12.5^{\circ}$&$1.19\pm0.04$&$-0.90\pm0.01$&$3.00$& $12.24\pm0.08$\\
     -     &     -    &  -  &  - &  -   &$0.08$&      -    &     -      &$1.38\pm0.04$&$-0.80\pm0.01$&$3.00$& $11.17\pm0.06$\\
     -     &     -    &  -  &  - &  -   &$0.10$&      -    &     -      &$1.60\pm0.08$&$-0.71\pm0.02$&$3.00$& $10.17\pm0.09$\\
\hline
$45^{\circ}$&$3^{\circ}$&$350$&$30$&$0.75$&$0.08$&$20^{\circ}$ &$12.5^{\circ}$&$0.41\pm0.02$&$-0.58\pm0.03$&$3.00$& $5.35\pm0.06$\\
     -    &     -    &  -  &  - &  -   &   -  &$45^{\circ}$ &     -       &$1.38\pm0.04$&$-0.80\pm0.01$&$3.00$& $11.17\pm0.06$\\
     -    &     -    &  -  &  - &  -   &   -  &$70^{\circ}$ &     -       &$2.72\pm0.09$&$-0.81\pm0.01$&$3.00$& $17.41\pm0.09$\\
\hline
$45^{\circ}$&$3^{\circ}$&$350$&$30$&$0.75$&$0.08$&$45^{\circ}$ &$10.0^{\circ}$&$1.70\pm0.12$&$-0.61\pm0.02$&$3.00$& $8.58\pm0.09$\\
     -    &     -    &  -  &  - &  -   &   -  &      -     &$12.5^{\circ}$&$1.38\pm0.04$&$-0.80\pm0.01$&$3.00$& $11.17\pm0.06$\\
     -    &     -    &  -  &  - &  -   &   -  &      -     &$15.0^{\circ}$&$1.21\pm0.03$&$-0.95\pm0.01$&$3.00$& $13.67\pm0.07$\\
\hline
\end{tabular}
\end{center}
\end{table*}

The density distribution of two patches can be defined on the pulsar
polar cap region and extended to high magnetosphere. For simplicity,
we model two density patches (see Fig.~\ref{fig:pmodel}) which are
symmetrically located about the meridional plane and peaked at
($\eta_m$, $\phi_m$) and ($\eta_m$, $-\phi_m$), respectively, with
Gaussian widths of $\sigma_\eta$ and $\sigma_\phi$ in the magnetic
polar and azimuthal directions, i.e.
\begin{equation}
n_{\rm e}(r,\eta,\phi)=n_{e0}(r/R_\star)^{-3}{\exp}[-\frac{(\eta-\eta_m)^2}{2\sigma_\eta^2}]{\exp}[-\frac{(\phi\pm\phi_m)^2}{2\sigma_\phi^2}].
\label{pmodel}
\end{equation}

The distribution of the emission intensity across height and the
rotation phase and the evolution of integrated profiles with frequency
for the curvature radiation model from particles of two density
patches are shown in Fig.~\ref{pimage} and \ref{pprofiles},
respectively. Our model calculations show that: 1) the leading
component is always slightly wider and weaker than the trailing one;
2) emission at a lower frequency comes from relatively higher and
wider region ($10-50 R_\star$ for $400$ MHz compared to $8-20 R_\star$
for $1.6$ GHz); 3) the low-frequency profiles become wider and are
shifted towards an early rotation phase; 4) the two components always
remain resolved even at very high frequencies, because two density
patches keep separated even down to the star surface.

\begin{figure}[tb]
    \centering
    \includegraphics[angle=0,width=0.44\textwidth]{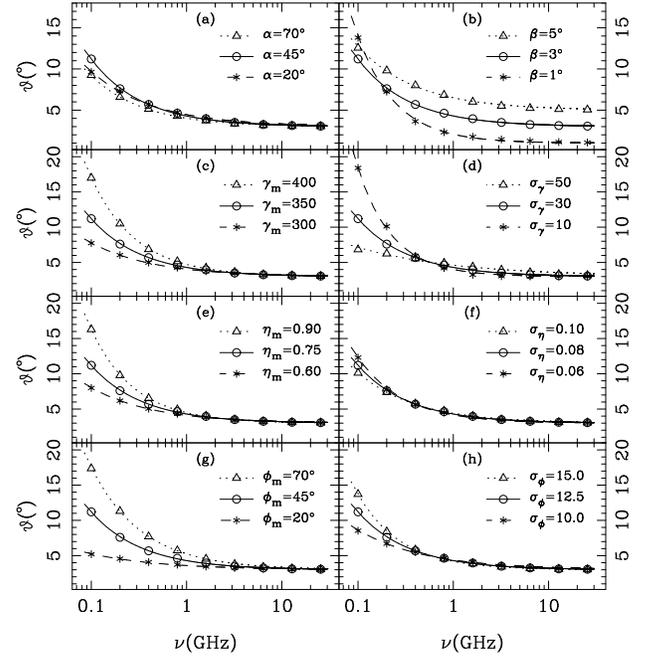}
    \caption{Same as Fig.~\ref{crelation} but for the model for two
     density patches. The model parameters are the same as those in
     Fig.~\ref{pimage}.}
    \label{prelation}
\end{figure}

From simulated profiles at a set of frequencies, we calculated the
beam radius and checked the frequency dependence of pulsar beam.  We
also checked if FDB curves are related to emission geometry ($\alpha$,
$\beta$), particle energy distribution ($\gamma$, $\sigma_\gamma$) and
patch geometry ($\eta_m$, $\sigma_\eta$, $\phi_m$ and
$\sigma_\phi$). The curves are shown in Fig.~\ref{prelation} and
fitted parameters are listed in Table~\ref{tpfdb}. The influence of
these parameters on the FDB curves are more or less similar to those
found from the conal density models. We noticed that $\phi_m$ has the
similar effect on the FDB curves as $\eta_m$, both of which determine
the separation of the two patches. Particles in two patches with a
larger $\sigma_\phi$ can produce the emission beams with a steeper FDB
curve.

\begin{figure}[htb]
\centering
\includegraphics[angle=-90, width=0.44\textwidth]{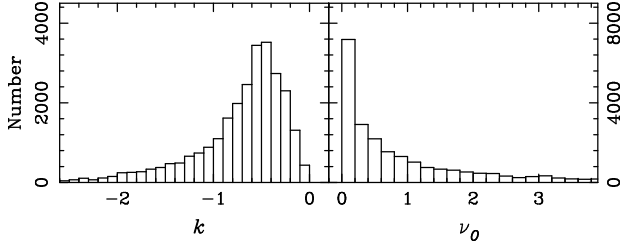}
\caption{Distributions of $k$ and $\nu_0$ from  29070 combinations of
various model parameter values of the two density patch model.}
\label{pknudis}
\end{figure}
\begin{figure}[htb]
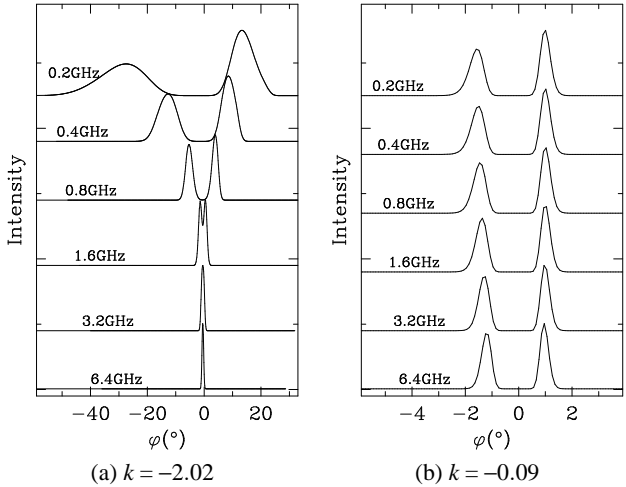

\centering
\begin{tabular}{cc}
\includegraphics[angle=0, width=0.21\textwidth] {P-profiles_largek.ps} &
\includegraphics[angle=0, width=0.21\textwidth] {P-profiles_smallk.ps} \\
{\small (a) $k=-2.02$} & {\small (b) $k=-0.09$}
\end{tabular}
\caption{Profile evolutions with frequency in the two density patch
  model for two extreme values of $k$, {\it Panel (a)} for $k=-2.02$
  and {\it Panel (b)} for $k=-0.09$.  Model parameters for panel (a)
  are $\alpha=45^\circ$, $\beta=5^\circ$, $\gamma_m=400$,
  $\sigma_\gamma=10$, $\eta_m=0.9$, $\sigma_\eta=0.1$,
  $\phi_m=65^\circ$ and $\sigma_\phi=15^\circ$, and for panel (b) are
  $\alpha=45^\circ$, $\beta=1^\circ$, $\gamma_m=200$,
  $\sigma_\gamma=90$, $\eta_m=0.3$, $\sigma_\eta=0.05$,
  $\phi_m=45^\circ$ and $\sigma_\phi=5^\circ$.}
\label{pprof2}
\end{figure}

To investigate the possible range of $k$ and $\nu_0$ in the two
density patch model, we have calculated the integrated profiles and
fitted the FDB curves for 29070 sets of models with different
combinations of model parameters of $\alpha = (5^\circ$, $45^\circ$,
$85^\circ$), $\beta = (1^\circ$, $5^\circ$, $9^\circ$), $\gamma_m =
(200$, $300$, $400$), $\sigma_\gamma = (10$, $50$, $90$), $\eta_m =
(0.1$, $0.3$, $0.5$, $0.7$, $0.9$), $\sigma_\eta = (0.05$, $0.1$,
$0.15$, $0.2$, $0.25$), $\phi_m = (25^\circ$, $45^\circ$,
$65^\circ$, $85^\circ$) and $\sigma_\phi = (5^\circ$, $15^\circ$,
$25^\circ$, $35^\circ$, $45^\circ$). The distributions of the
power-law indices $k$ and the characteristic frequency $\nu_0$ are
shown in Fig.~\ref{pknudis}.  We noticed that $k$ values are in the
range between $-0.1$ and $-2.5$. Two components are always resolved,
except for the extreme case shown in Fig.~\ref{pprof2}(a) in which
emission is radiated by particles with a narrower energy
distribution ($\sigma_\gamma = 10$) and peaking at a larger energy
($\gamma_m = 400$) at very high $\eta_m = 0.9$ and $\phi_m = 65$.

\begin{figure}[tb]
    \centering
    \includegraphics[angle=-90,width=0.3\textwidth]{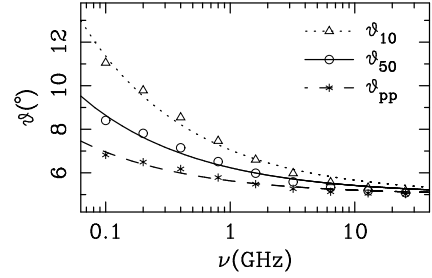}
    \caption{Beam radii ($\vartheta_{10}$) calculated from the 10\%
      profile widths ($W_{10}$) at a series of frequencies are
      compared with beam radii ($\vartheta_{50}$) calculated from the
      50\% profile widths ($W_{50}$) and beam radii ($\vartheta_{pp}$)
      calculated from component separations ($W_{pp}$). Parameters for
      this example model are $\gamma_m=200$, $\sigma_\gamma=50$,
      $\eta_m=0.7$, $\sigma_\eta=0.2$, $\phi_m=65^\circ$ and
      $\sigma_\phi=15^\circ$, with $P=1s$, $\alpha=45^\circ$ and
      $\beta=5^\circ$.}
    \label{p1050pp}
\end{figure}

\begin{figure}[tb]
    \centering
    \includegraphics[angle=-90,width=0.45\textwidth]{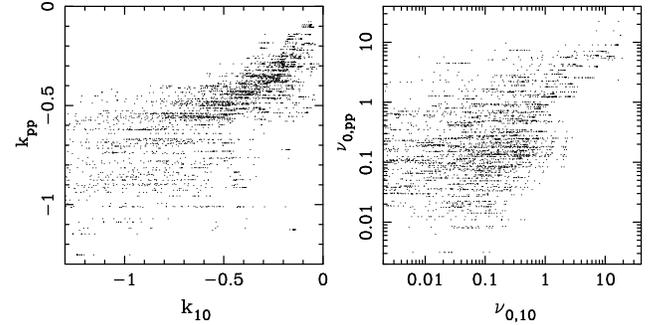}
    \caption{Comparison of $k$ and $\nu_0$ values calculated from
      $\vartheta_{10}$ and $\vartheta_{pp}$}
    \label{kknunu}
\end{figure}

In above beam calculations, we used the profile width $W_{10}$ at
10\% of pulse peak. If the separations $W_{pp}$ of two components
are used to calculated the beam radii, the beam sizes are obviously
smaller than those calculated from the 50\% and 10\% widths
(Fig.~\ref{p1050pp}).  We compared in Fig.~\ref{kknunu} the $k$ and
$\nu_0$ values calculated from $\vartheta_{10}$ (written as $k_{10}$
and $\nu_{0,10}$) with the $k$ and $\nu_0$ values from
$\vartheta_{pp}$ (written as $k_{pp}$ and $\nu_{0,pp}$), and found
that they are correlated and scatter at small $\nu_0$ or large $k$
values (due to error-bar as expected).

In summary, the curvature radiation from particles in two density patches
can not only give a proper range of $k$ values, but also keeps the two
components resolved in general even at very high frequencies.

\section{Discussions}

In the previous section, we conclude that the FDB curves with two
resolved components can be preferably explained by curvature
radiations of particles in two density patches.  Here we discuss
influences on the FDB curves by the radial decay of particle energy
and by other possible energy distribution of particles, and discuss
the possible FDB curve flattening at low frequencies. Finally we will
look at the real FDB data and make a model.

\subsection{Radial decay of particle energy and the lowest emission height}

As shown in Fig.~\ref{crelation} and \ref{prelation}, the values of
beam radii at the infinite-frequency, $\vartheta_0$, calculated from
models are always very close to those of $\beta$, no matter what
model parameters are in putted. However, both on observational
ground (Table.~\ref{psr-para}) and theoretical expectations,
$\vartheta_0$ is larger than $\beta$.

In simulations above we assume that particles in the magnetosphere
have a Gaussian energy distribution with a peak Lorentz factor
$\gamma_m$ and a spread width $\sigma_\gamma$ along the whole
trajectory. In reality, the particle energy distribution may be much
more complicated due to the Compton loss \citep{zqh97} or other
processes. It is reasonable to believe that particle energy decreases
when they flow out along the field lines in the inner magnetosphere.

In the vacuum-gap model \citep{rs75}, electron-positron pairs are
produced by the avalanche process in the pulsar polar cap. The maximum
Lorentz factor of the primary particles after accelerating across the
gap is $\gamma_{\max}=1.2\times10^7B_{\star 12}h_4^2/P_{\rm 1s}$, with
$B_{\star 12}=B_\star/10^{12}$\,G the surface magnetic field,
$h_4=h/10^4\,$cm the gap height and $P_{\rm 1s}=P/{\rm 1s}$. Pair
cascade occurs within a few stellar radii near the surface, which
generates the secondary particles with much lower Lorentz factors of a
few hundred. Detailed numerical simulations on the pair cascade
process were carried out by \citet{ml10}, in which they found that for
various initial parameters (such as surface magnetic field, period and
primary energy), the pair cascade process maybe finishes at a height
of several star-radius, from less than $2R_\star$ to more than
$10R_\star$.
Here we naively assume a toy model to describe the rapid decay of
particle energy during the pair cascade process, which reads
\begin{equation}
\gamma_m=\gamma_{m0}\exp(-q\frac{r-R_{\star}}{R_{\star}})+\gamma_s,
\label{gamma_m}
\end{equation}
here $\gamma_m$ is the peak Lorentz factor of the Gaussian
distribution we used above, $\gamma_{m0}$ and $\gamma_s$ are the
corresponding values of the primary particles and the final secondary
particles before and after the pair cascade, respectively. The factor
$q$ (less than 1) determines how fast the energy decreases with
height for the cascade. This energy model has been used by
\citet{qlz+01} and \citet{zqh+07} to simulate the inverse-Compton
scattering emission, where they ignored the energy term $\gamma_s$. To
use the above energy model in our simulation for the model for two
density patches, one should substitute $\gamma_m$ of
Eq.~(\ref{gamma_m}) into the Gaussian energy distribution
Eq.~(\ref{gammadis}). Note that the number of the particles $n_{\rm
  e0}$ in Eq.~(\ref{pmodel}) should also be replaced by $n_{\rm
  e0}\gamma_s/\gamma_m$ to keep the conservation of total energy.

\begin{figure}[tb]
    \centering
    \includegraphics[angle=0, width=0.3\textwidth]{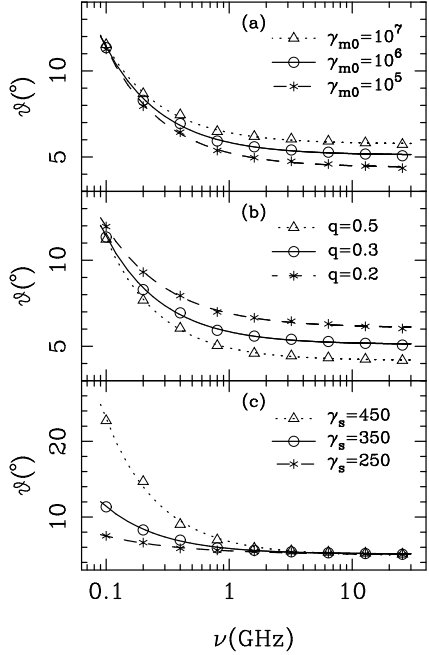}
    \caption{FDB curves for the curvature radiation model by particles
      in two density patches with an energy distribution of
      Eq.~(\ref{gammadis}) and Eq.~(\ref{gamma_m}). The default model
      parameters are the same as those in Fig~\ref{pimage}
      (e.g. $\beta=3^{\circ}$) but also with $\gamma_{m0}=10^6$,
      $q=0.3$ and $\gamma_s=350$. }
\label{gmdecfdb}
\end{figure}

\begin{table}[tb]
\small
\begin{center}
\scriptsize
\caption{Fitted parameters of FDB curves in Fig.~\ref{gmdecfdb}}
\label{gmdecfdbt}
\begin{tabular}{ccc|ccc}
\hline \hline
 $\gamma_{m0}$&$q$&$\gamma_s$&$\nu_0 (\rm GHz)$&$k$& $\vartheta_0 (^{\circ})$ \\
 \hline
 $10^5$  &  $0.3$  & $350$ &$0.84\pm0.05$&$-0.91\pm0.03$&$4.39\pm0.05$\\
 $10^6$  &     -   &   -   &$0.72\pm0.05$&$-0.93\pm0.03$&$5.11\pm0.05$\\
 $10^7$  &     -   &   -   &$0.64\pm0.05$&$-0.94\pm0.04$&$5.76\pm0.05$\\
 \hline
 $10^6$  &  $0.2$  & $350$ &$0.82\pm0.06$&$-0.84\pm0.03$&$6.07\pm0.05$\\
    -    &  $0.3$  &   -   &$0.72\pm0.05$&$-0.93\pm0.03$&$5.11\pm0.05$\\
    -    &  $0.5$  &   -   &$0.72\pm0.03$&$-0.99\pm0.02$&$4.18\pm0.03$\\
\hline
 $10^6$  &  $0.3$  & $250$ &$0.35\pm0.04$&$-0.71\pm0.06$&$5.06\pm0.05$\\
    -    &    -    & $350$ &$0.72\pm0.05$&$-0.93\pm0.03$&$5.11\pm0.05$\\
    -    &    -    & $450$ &$1.85\pm0.30$&$-0.99\pm0.06$&$4.93\pm0.22$\\
\hline
\end{tabular}
\end{center}
\end{table}

Example FDB curves from the curvature radiation model by particles
in two density patches with an energy distribution described by
Eq.~(\ref{gammadis}) and Eq.~(\ref{gamma_m}) are demonstrated in
Fig.~\ref{gmdecfdb}, and the fitted parameters are listed in
Table.~\ref{gmdecfdbt}. We found that a larger primary Lorentz
factor $\gamma_{m0}$ or a smaller $q$ lead to larger beam radius
$\vartheta_0$ at very high frequencies. A larger $\gamma_s$ leads
to a steeper FDB curve.

In these model the values of $\vartheta_0$ are always larger than
those of $\beta$, which means that there exists the lowest radio
emission height $r_{\rm low}=R_{\rm lc}\sin^2(2\vartheta_0/3)$, which
is $> R_{\rm lc}\sin^2(2\beta/3)$. For a small value of $\vartheta_0$,
the corresponding beam radii at the infinite frequency can be written
as $\vartheta_0\simeq3/2\sqrt{r_{\rm low}/R_{\rm lc}}$. Now, what is
the value of $r_{\rm low}$?

\begin{figure}[tb]
    \centering
    \includegraphics[angle=0, width=0.45\textwidth]{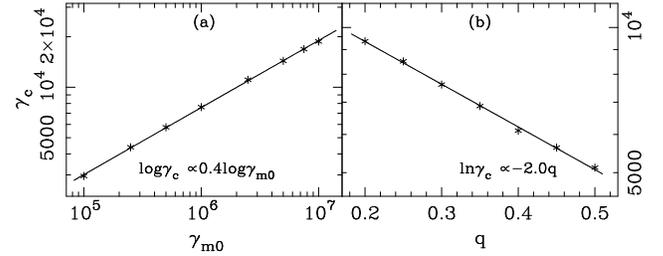}
    \caption{Characteristic Lorentz factor $\gamma_c$ at lowest
      radio emission height well related to $\gamma_{m0}$ and
      $q$. Symbols $*$ come from simulation data as in
      Table~\ref{gmdecfdbt} and lines are the fitting.}
    \label{fig:gammac}
\end{figure}

According to Eq.~(\ref{gamma_m}), at the lowest radio emission height
$r_{\rm low}$, there exists a characteristic Lorentz factor
$\gamma_c$. We noticed that $r_{\rm low}$ and $\gamma_c$ are related
with $\gamma_{m0}$ and $q$ by:
\begin{equation}
\frac{r_{\rm
low}}{R_\star}=1-\frac{1}{q}\ln\frac{\gamma_c}{\gamma_{m0}}.
\label{eq:rlow}
\end{equation}
From simulations similarly as presented Table~\ref{gmdecfdbt}, we can
get a set of $r_{\rm low}$ from $\vartheta_0$ for various
$\gamma_{m0}$ and $q$, and then we got $\gamma_c$ using
Eq.~(\ref{eq:rlow}).  We find that $\gamma_c$ are also very well
related to $\gamma_{m0}$ and $q$, as shown in Fig.~\ref{fig:gammac},
by
\begin{equation}
\gamma_c=55.0\gamma_{m0}^{0.4} \exp{(-2q)}.
\label{eq:gammac}
\end{equation}
Substituting into Eq.~(\ref{eq:rlow}), we find
\begin{equation}
{r_{\rm low}}= {R_\star}(3+\frac{1.38\log\gamma_{m0}-4}{q}).
\label{eq:rlow2}
\end{equation}

\begin{figure}[tb]
    \centering
    \includegraphics[angle=0, width=0.21\textwidth]{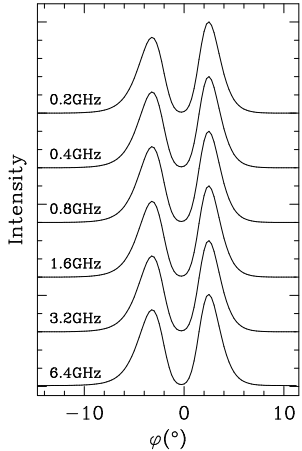}
    \caption{Same as Fig.~\ref{pprofiles} but a power-law energy
      distribution with index $s=-2$. 
      Other model parameters are $\gamma_{s,\rm min}=150$ and
      $\gamma_{s,\rm max}=2000$, $\gamma_m=350$, $\sigma_\gamma=30$,
      $\eta_m=0.75$, $\sigma_\eta=0.08$, $\phi_m=45^{\circ}$ and
      $\sigma_\phi=12.5^{\circ}$, with $P=1$s, $\alpha=45^{\circ}$ and
      $\beta=3^{\circ}$.}
\label{pl-profiles}
\end{figure}

\subsection{On the energy distribution of particles}

At any height, the secondary particles in the magnetosphere for the
curvature radio emission were assumed to have an energy distribution
in the Gaussian function. It is possible that the particles follow
other energy distribution functions. For example, according to
\citet{aro81a}, the multi-component plasma in pulsar magnetosphere
contains secondary pairs, a high-energy plasma `tail', and some
primary particles. The joint energy distribution of secondary
particles and high-energy `tail' can be described by a power-law
with two cutoffs at its two ends, which reads as,
\begin{equation}
n_{\rm e}(\gamma)\sim \gamma^s,\,\,\,\, \gamma_{s,\rm min}<\gamma<\gamma_{s,\rm max},
\label{gammapl}
\end{equation}
where $s$ is the power-law index.
The curvature radiation from particles in two density patches with
such an energy distribution produces the integrated profiles at a
series of frequencies with a similar shape and a constant pulse width
between 80 MHz and 40 GHz, as shown in Fig.~\ref{pl-profiles}. We can
not find the profile differences for a range of $s$ values, e.g. a
difference less than $2\%$ for the pulse width values between $s=-2$
and $s=-3$. We also found that the values of $\gamma_{s,\rm min}$ and
$\gamma_{s,\rm max}$ determine the frequency range for the constant
profile width.

\begin{figure}[tb]
    \centering
    \includegraphics[angle=-90,width=0.3\textwidth]{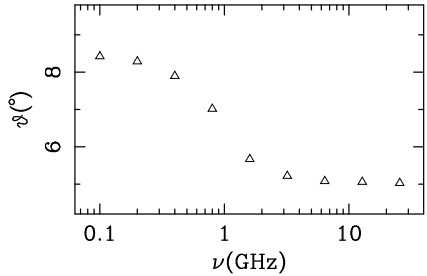}
    \caption{Example of Low frequency flattening the FDB curve. The
      model parameters are $\gamma_m=400$, $\sigma_\gamma=10$,
      $\eta_m=0.7$, $\sigma_\eta=0.1$, $\phi_m=45^\circ$ and
      $\sigma_\phi=5^\circ$ with $P=1s$, $\alpha=45^\circ$ and
      $\beta=5^\circ$.}
    \label{pflatten}
\end{figure}

\subsection{Possible low frequency flattening of the FDB curves}

During the model calculations, we noticed that the FDB curves tend to
be flattening at low frequencies. For some input parameters of the two
density patch model, the effect is very obvious (Fig.~\ref{pflatten}),
especially when Lorentz factors are large ($\gamma_m>400$) and/or the
energy distribution is broad ($\sigma_\gamma>80$).

This is understandable. Because of the dipole bending of magnetic
field lines, as pulsar rotates, a fixed sight-line cut across the
two density patches (see Fig.~\ref{fig:pmodel}) with a clear lower
limit and upper limit of height and a clear left and right limit of
the rotation phase.  Very low frequency emissions tend to come from
the high regions of two patches in such a limited volume. In that
high regions near the upper bound, the lower frequency emission
comes from particles of the lower energy, but the pulse-width or
beam-width of the emission from two patches is determined by the
geometry. This happens to very low-frequency, depending on the
$\gamma_m$.  The low-frequency flattening of FDB curves are often
seen in the models when patches are small (small $\sigma_\phi$), or
energy of particles are very high and/or the energy distribution is
broad (large $\gamma_m$ and $\sigma_\gamma$).

\begin{figure}[tb]
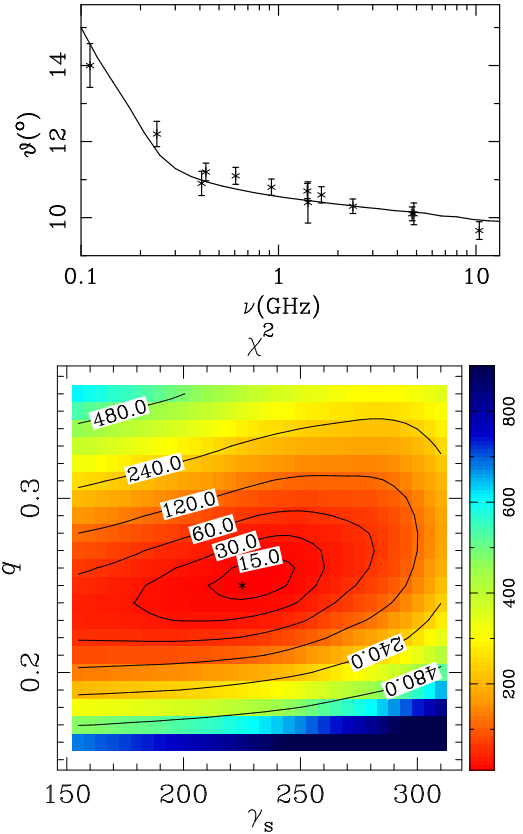

    \centering
    \includegraphics[angle=0, width=0.35\textwidth]{comp2020.ps}
    \includegraphics[angle=0, width=0.37\textwidth]{chisquare-dis.ps}
    \caption{{\it Upper panel:} An example of real data and
      model-fitting. The observed data of PSR B2020+28 are taken from
      \citet{mr02a}. The solid line is the best model calculated from
      the curvature radiation of particles in two density patches with
      parameters $\gamma_{m0}=10^6$, $\gamma_s=225$,
      $\sigma_\gamma=15$, $q=0.25$, $\eta_m=0.75$, $\sigma_\eta=0.08$,
      $\phi_m=45^{\circ}$ and $\sigma_\phi=12.5^{\circ}$. The
      geometrical parameters $\alpha$ and $\beta$ and the period of
      this pulsar can be found from Table 1. {\it Lower panel:} The
      $\chi^2$ distribution for a pair of the most influential
      parameters, $\gamma_s$ and $q$, for the fitting. The best values
      are $\gamma_s=225$ and $q=0.25$.}
\label{compa}
\end{figure}

\subsection{Observed FDB curve and a model}

We have worked on a huge number of models for the radio emission
intensity distribution across the height and rotation phase and
integrated profiles. We now look at data of a real cone-dominant
pulsar, PSR B2020+28, and make a model. The data are taken from
\citep{mr02a}. To keep the two profile components separated even at
very high frequencies, we have to choose a model of the curvature
radiations of particles in two density patches and consider the energy
distribution of particles as described by Eq.~(\ref{gammadis}) and
Eq.~(\ref{gamma_m}). As shown in Figure~\ref{compa}, the model can fit
data nicely with a local minimum of $\chi^2=10.24$ in the parameter
space of $\gamma_s$ and $q$. The model is obtained with sparsely
separated parameter grids, which may be improved by a global fitting
for 8 parameters but that is very computational expensive.

\section{Summary and outlook}

We studied the frequency dependence of beam radius for cone-dominant
pulsars by using the curvature radiation mechanism.

For the simplest case in which pulsar radio emission is generated by
curvature radiation of relativistic particles with a single $\gamma$
and streaming along fixed open field-lines, we obtain the analytic
formula in Eq.~(\ref{appro}) for the frequency dependence. It has the
power-law index $k=-1$, and $\vartheta_0=0$, both are different from
observations.

Considering various density distribution and energy distribution of
particles, we numerically calculated the intensity of curvature
radiation from all the possible emission heights in the open field
line region, and get integrated mean pulse profiles from which the FDB
curves are derived. We found that the power-law index $k$ and
$\vartheta_0$ are strongly influenced by the density distribution and
energy distribution of particles in the open field-line region.

When particles are uniformly distributed in whole open field lines,
the curvature radiation can produce only one profile component,
which can not explain the two resolved components
of cone-dominant pulsars.

When particles are distributed in a conal area, the low-frequency
emission comes from very high regions so that profile is very wide.
Profiles at high frequencies cannot have two separated components,
because emission comes from low heights where the sight line always
continually cuts the edge of density cone.

To keep two profile components separated or resolved, the particles
must come from two density patches. Model calculations from a huge
amount of parameter combinations show that the $k$ values of the FDB
curves are in the range of $-0.1$ to $-2.5$, in agreement with the
observations. Because of the limited upper boundary of the density
patches cut by a given sight line, the FDB curves tends to be
flattened at low frequencies, especially for models with a large
Lorentz factor and/or broad particle energy distribution.

The geometry of density distribution of particles and the energy
distribution of particles in the model certainly affect the FDB
curves.  For a Gaussian energy distribution, high energy particles
with large $\gamma_m$ will produce a wide beam at low
frequency. Particles in a narrow energy distribution with a small
$\sigma_\gamma$ produce a steep FDB curve. The radial decay of
particle energy is important to get a reasonable $\vartheta_0$ in
models.

Through model calculations of pulsar beam and profile evolution by
using the curvature radiation mechanism, we have clarified some
important issues on particle density and energy distributions.
However, in our simulations, we considered only the simple case that
particles are emitting radio waves with the characteristic frequency
for curvature radiation determined by the Lorentz factor. The real
curvature emission of any particles has an energy spectrum, which is
a power-law with an index $1/3$ at the low frequency limit (much
less than the characteristic frequency) and becomes exponential at
high frequencies. This may significantly affect the emission regions
and modify the FDB curves, which we should investigate in future.
Another important fact is that the real pulsar radio emission is
believed to be highly coherent, while in this paper we simply treat
the coherence by assuming that the emission comes from point-like
huge charge. Moreover, other emission mechanisms may also work in
pulsar magnetosphere, such as inverse Compton scattering, plasma
oscillation, cyclotron instability and Coulomb bremsstrahlung
emission, etc, which are not considered in this paper and may lead
to their own FDB curves.

\acknowledgments The authors thank Professor Chou Chih Kang for
helpful discussions and the referee for constructive comments. This
work has been supported by the National Natural Science Foundation of
China (11003023 and 10833003).


\end{document}